\documentclass[14pt]{article}
\usepackage{epsfig}
\usepackage[english]{babel}
\usepackage[T1]{fontenc}
\usepackage{graphicx}
\usepackage{amsmath, amssymb, latexsym, amsopn}

\begin{document}

\title{Conditions for spontaneous homogenization of the Universe}
\author{Krzysztof Bolejko\thanks{bolejko@camk.edu.pl}\\
Steward Observatory, University of Arizona, Tucson, AZ 85721, USA \\
and Nicolaus Copernicus Astronomical Center, Bartycka 18, 00-716 Warszawa, Poland \\
\and
William R. Stoeger \\
Vatican Observatory Research Group, University of Arizona, Tucson, AZ 85721, USA}
\maketitle

\begin{abstract}
The present-day Universe appears to be homogeneous on very large scales.
Yet when the casual structure of the early Universe is considered, it becomes apparent that the early Universe must have been highly inhomogeneous. 
The current paradigm attempts to answer this problem by postulating the inflation mechanism
However, inflation in order to start requires a homogeneous patch of at least the horizon size.
This paper examines if dynamical processes of the early Universe could lead to homogenization.
In the past similar studies seem to imply that the set of initial conditions that leads to homogenization is of measure zero. This essay proves contrary:
a set of initial conditions for spontaneous homogenization of cosmological models
can form a set of non-zero measure.
\end{abstract}

\vspace{1.9cm}

\begin{center}
{\large {\em Essay written for the Gravity Research Foundation 2010 Awards \\
for Essays on Gravitation}}
\end{center}

\newpage

The present-day Universe appears to be homogeneous on very large scales. The main argument comes from the isotropy of the cosmic microwave background radiation (CMB). The Ehlers-Geren-Sachs (EGS) theorem \cite{EGS} and `almost EGS theorem' \cite{SME} imply that if anisotropies in the CMB are small for all fundamental observers then locally the Universe is almost spatially homogeneous and isotropic. In addition, if there were large inhomogeneities in the Universe they would manifest themselves in CMB temperature fluctuations via the Ree-Sciama effect \cite{RS68,IS06,IS07}. Moreover, the enormous success of the homogeneous 
Friedmann--Lema\^itre--Robertson--Walker (FLRW) models in describing cosmological observations seems to provide further evidence for the large scale homogeneity of our Universe --- nearly all sets of cosmological observations are successfully analyzed within the framework of homogeneous models, including supernova observations \cite{sn1,sn2}, baryon acoustic oscillations \cite{bao} and the CMB \cite{cmb}.

On the other hand, when the casual structure of the early Universe is considered, it becomes apparent that the early Universe must have been highly inhomogeneous \cite{W84,E06}. This raises a key question: {\em how did the chaotic behavior expected in the very early Universe come to generate the Cosmos that on large scales looks so homogeneous\/}?

The current paradigm attempts to answer this question by postulating the inflation mechanism \cite{L90,VM05}. Inflation not only appears to solve the horizon problem but also other problems of the standard cosmology like the origin of matter fluctuations (during inflation quantum fluctuations are enhanced and they become a source of primordial fluctuations), magnetic monopole (while the particle physics predicts existence of magnetic monopoles or other exotic particles, these have not been observed as during  inflation they were pushed outside the horizon), flatness problem (since the inflating region increases its size by 
factor $\sim10^{33}$ the effective curvature decreases by $\sim10^{-66}$), and many others. While alternative theorems such as cosmic stings or bounce cosmologies can deal with some of the problems they do not solve all the problems in such a graceful and simple way as inflation does \cite{L08}.

However, after closer investigation it becomes apparent that inflation does not solve the problem of homogenization of the Universe. Rather, it explains how it is possible that regions which appear presently causally disconnected 
were casually connected at some very early stage of the evolution of the Universe. This is because the mechanism of inflation relies on explicit models that are homogeneous from the very beginning. The studies of inflation within inhomogeneous backgrounds proved that in order to start, inflation requires a homogeneous patch of at least the horizon size
\cite{GP89,GP90,GP92,GS92}.

Ironically, inflation which had originally been postulated to explain homogeneity, now relies on the assumption that the early Universe was homogeneous. 
This is what the lay people call a vicious circle. One may try to avoid the problem by pointing that there is always a probability that a homogeneous patch (even of a horizon size) existed in the Universe, or even by invoking the Anthropic Principle arguing that if inflation could not start then the galaxies and life could not form. However, the above arguments do not seem to be satisfactory --- using similar arguments one could try to argue that the Universe was from the very beginning  homogeneous and the initial perturbations were encrypted in it and then there would be no need for inflation at all.
Given that we can study only one Universe, the problem of initial conditions of the Universe seems beyond the reach of modern science \cite{E06}.
Also, our current knowledge about matter and its properties
under extreme conditions seems rather insufficient to study conditions of the early Universe which most probably should be described by the theory of quantum gravity.
Nevertheless, the problem of homogenization should also be studied using classical methods. An elegant solution would be to prove some dynamical scenario (other than sheer luck) in which starting from arbitrary initial conditions the Universe would undergo spontaneous homogenization. On the other hand, if one can show that no dynamical process can homogenize the Universe before the start of inflation this would be a strong suggestion that initial conditions had to be tuned in order to achieve homogenization, or that there must be some other process that can explain  current cosmological problems without inflation.

The problem of dynamical homogenization of the Universe is not new. In 1968 Misner \cite{M68} proposed that the Universe started off in a chaotic state and that dissipation  processes damped out inhomogeneities leading to a homogeneous state. Misner considered the Bianchi type I solution with viscosity. Among homogeneous and anisotropic models
the Bianchi I models are the simplest and in the case of vanishing shear they reduce to the parabolic FLRW solution, and they are characterized by the following metric

\begin{equation} {\rm d} s^2 =  - {\rm d} t^2 + X^2(t) {\rm d}  x^2 + Y^2(t) {\rm d} y^2 + Z^2(t) {\rm d} z^2. \end{equation}
It can be shown \cite{HE73} that the function $S^3(t) = XYZ$ evolves as

\begin{equation} S = \frac{2}{3 t} \frac{t+ \Sigma/2}{t+\Sigma},\end{equation}
where $\Sigma$ is a constant. As seen if $t \gg \Sigma$, anisotropies decrease ---
consider a simple case of $Y=Z$ then the shear is equal to

\begin{equation} \sigma^2 = \frac{1}{3 t^2} \left( \frac{\Sigma}{t+\Sigma} \right)^2. \end{equation}
For $t\gg \Sigma$, $\sigma^2 \sim a^{-6}$, where $a(t)$ is the scale factor of the 
FLRW model, and $S(t) \xrightarrow[t\gg\Sigma]{} a(t)$.
If instead of dust a viscous fluid is considered then at late times, 
as shown  by Ellis \cite{E71}

\begin{equation} \sigma^2 \sim \frac{\exp[-2\lambda t] } {a^6}, \end{equation}
where
$\lambda$ is the viscosity coefficient.
This simple example shows that viscosity can induce an exponential decay of the shear
and therefore be a source of isotropization.

However, Collins \& Hawking \cite{CH73} showed that among homogeneous and anisotropic models
those that approach homogeneous and isotropic state are of measure zero. They also suggested that within inhomogeneous framework homogenization  seems unlikely. On the other hand Bonnor and Tomimura \cite{B74,BT76} showed that within a wide class of initial conditions  inhomogeneous models (e.g. Lema\^itre-Tolman and Szekeres geometries) which evolve into the homogeneous FLRW models can exist. In other words, there exist inhomogeneous solutions with only decaying modes.
As later confirmed by Silk \cite{S77} and Pleba\'nski \& Krasi\'nski \cite{PK06} such models are characterized by being free of curvature perturbation. 
One may think that models with only decaying modes form a set of measure zero,
since one of arbitrary functions defining such models must be of specific form.
However, there are some cases when perturbations are described {\em only}
by the decaying modes. For example let us consider a `comformal-Newtonian' metric

\begin{equation} {\rm d} s^2 =  
a^2 \left[ (1+\Phi) {\rm d} \eta^2 -(1-2\Psi) \delta_{ij} {\rm d} x^i {\rm d} d^j \right],
\end{equation}
For the ultra-relativistic matter or radiation-dominated Universe ($p/\rho = 1/3$)
the adiabatic perturbations evolves as \cite{VM05}

\begin{equation} 
\Phi_k = \frac{1}{x^2} \left[ C_1 \left(\frac{\sin x}{x} - \cos x \right)
+ C_2 \left( \frac{\cos x}{x} + \sin x \right) \right],
\end{equation}
where $x = k\eta/\sqrt{3}$, $k$ is a wavenumber, and $\eta$ is a conformal time 
(${\rm d} t = a {\rm d} \eta$). Since $ \Phi \simeq  -2 {\delta \rho}/{\rho}$ 
then the adiabatic perturbations of the ultra-relativistic matter are always described by decaying modes. Since at high temperatures matter should be ultra-relativistic,
this shows that the early Universe could provide suitable
conditions for homogenization.

So far we have considered only simple examples, now
we consider whether in more general inhomogeneous case the set of initial conditions that lead to homogenization can be of non-zero measure. Let us assume that there is a unique timelike vector field
$u^{a}$ which can be associated with the average  velocity of matter \cite{E71},
then the tensor

\begin{equation}  h^{a}{}_{b} = \delta^a{}_b + u^a u_b, \end{equation}
projects tensors into the surface orthogonal to $u^a$.
Further the velocity field can be decomposed \cite{E71}

\begin{equation} 
u_{a;b} = \omega_{a b} + \sigma_{a b}
+ \frac{1}{3} h_{a b} \theta - \dot{u}_a u_b,
\end{equation} 
into shear ($\sigma_{a b})$, rotation ($\omega_{a b}$), expansion
($\theta$)  and acceleration ($\dot{u}_a = u_{a;b} u^b$).
The homogeneous solutions form a set of vanishing rotation, shear and acceleration.
Therefore, a potential process of homogenization of the universe must
result in reducing the amplitude of 
$\omega_{a b}$, $\sigma_{a b}$,
and $\dot{u}_a$ to some negligible values.
Let us focus on the shear evolution which is given by \cite{E71}

\begin{equation} \label{shearev}
h_a{}^f h_b{}^g \dot{\sigma}_{fg}
- h_a{}^f h_b{}^g \dot{u}_{(f;g)}
- \dot{u}_a \dot{u}_b + \omega_a \omega_b
+ \sigma_{a}{}^f \sigma_{fb}
+ \frac{2}{3} \theta \sigma_{ab}
+ h_{ab} ( - \frac{1}{3} \omega^2 - \frac{2}{3} \sigma^2 + \frac{1}{3} \dot{u}^c{}_c
) - \frac{1}{2} \pi_{ab} + E_{ab} = 0.
\end{equation}
Since the propagation equation for $\omega_{a b}$ and $\dot{u}_a$
are of similar complexity it is hard to say, without detailed numerical calculations
how the Universe can homogenize itself.
However, if the Universe is expansion-dominated, i.e the term 
$\theta \sigma_{ab}$ dominates in the above equation, then
the shear decays. In the early Universe
the expansion rate is indeed of high amplitude, but so
the other terms can be. 
To build up our intuition about dominance of $\theta \sigma_{ab}$ let us consider
a few simple examples.
For simplicity we will consider space-times that are 
locally rotationally symmetric (LRS).
In such a case at each point there exists  a unique preferred spatial direction
which constitutes a local axis of symmetry \cite{vEE96}.
Therefore, there exists a spacelike vector field such that

\begin{equation}
e^i u_i = 0, \quad e_i e^i = 1 \Rightarrow e^i e_{i;j} = 0.
\end{equation}
Using this vector field one can define a trace-free symmetric tensor

\begin{equation}
e_{(ij)} = \frac{1}{2} \left( 3 e_i e_j - h_{ij} \right).
\end{equation}
Due to the LRS property all covariantly defined spacelike trace-free symmetric tensors which are
orthogonal to $u^{a}$ have to be proportional to $e_{ij}$. Therefore,
the shear and the electric part of the Weyl tensors can be written as follows

\begin{equation}
\sigma_{(ij)} = \frac{2}{\sqrt{3}} \sigma e_{ij}, \quad  E_{(ij)} = \frac{2}{\sqrt{3}} E e_{ij},
\end{equation}
where $\sigma = \frac{1}{2} \sigma^i{}_j \sigma^j{}_i$ and $E =  \frac{1}{2} E^i{}_j E^j{}_i$. 

Let us now consider the Lema\^itre--Tolman model ($p=0=\omega = \dot{u}$).
The metric of the Lema\^itre--Tolman  model is \cite{L33,T34}

\begin{equation}
{\rm d} s^2 = - {\rm d} t^2 + \frac {{R,_r}^2}{1 + 2E(r)}{\rm d} r^2 +
R^2(t,r)({\rm d}\vartheta^2 + \sin^2\vartheta{\rm d}\varphi^2),
\end{equation}
where $E(r)$ is an arbitrary function, 
$R,_r = \partial R/ \partial r$, and
$R(t,r)$ obeys
 \begin{equation}\label{vel}
{R,_t}^2 = 2E + \frac{2M}{R} + \frac{\Lambda}{3} R^2,
 \end{equation}
where ${R,_t} = R_{;a} u^{a}$, $M(r)$ is another arbitrary function and $\Lambda$ is the cosmological constant. Below it will be assumed that 
$\Lambda =0$, first because the cosmological constant is negligible 
in the early universe, and second because we are interested in decelerating
cosmology (in contrast to accelerated phase of inflation).
The interpretation of the functions $M(r)$ and $E(r)$ is as follows:
$M$ is the gravitational mass contained within the
comoving spherical shell at any given $r$, while $E$ is the energy per unit
mass of the particles on that shell. Moreover, $E$ determines the space
curvature at each $r$-value.
For the Lema\^itre--Tolman model eq. (\ref{shearev}) reduces to

\begin{equation}\label{spLT}
\dot{\sigma} 
=  - \frac{1}{\sqrt{3}} \sigma^2 
- \frac{2}{3} \theta \sigma - \frac{1}{2 \sqrt{3}} (\rho -\overline{\rho}),
 \end{equation}
where $\rho=\frac{2M'}{R^2 R'}$ and $\overline{\rho} = \frac{6M}{R^3}$ (in the FLRW limit $\rho = \overline{\rho}$).
Since $\theta = 2T +P$ and $\sigma = \frac{1}{\sqrt{3}} (P-T)$ ---
where $P = \dot{R}'/R'$ is the parallel Hubble parameter,
and $T = \dot{R}/{R}$ is the tangential Hubble parameter ---
the second term on the rhs dominates over the first one.
On the other hand the last term is

\begin{equation}  \rho - \overline{\rho} = 2\sqrt{3} T \sigma  - \frac{2E'}{RR'} + \frac{4E}{R^2}.\end{equation}
It is therefore easy to notice that the parabolic 
Lema\^itre--Tolman model ($E=0$) undergoes homogenization.
The same is true if the second and third terms of the rhs are much smaller than $2T(P-T)$.
Let us consider, however, a stronger condition

\begin{equation}\label{conditionLT}
\frac{E'}{RR'} \ll \rho, \quad \frac{E}{R^2} \ll \rho.
\end{equation}
The above condition has an intuitive simple, geometric interpretation:
it is  equivalent to ${^3\mathcal{R}} \ll {\mathcal{R}}$
(where ${^3\mathcal{R}}$ is the Ricci scalar of 3-space, and
$\mathcal{R}$ is the Ricci scalar of space-time). Therefore, if only 
(\ref{conditionLT}) holds universe homogenize itself.

Surprisingly, eqs. (\ref{spLT}) and (\ref{conditionLT})
are literally the same for the quasi-spherical Szekeres model \cite{S75} (which
do not have any Killing vectors) and also for its generalization in
the  Roy-Singh model \cite{RS82,AK97} (where bulk and 
shear viscosity are of such a special form so that $\dot{u}^a = 0$).
For the Roy-Singh model $E_{ab} = \frac{1}{3} (\rho -\overline{\rho}) e_{ab} + \frac{1}{2} \pi_{ab}$, and thus the viscosity cancels out in (\ref{shearev}).
This shows that the spherically symmetric dust solutions 
of the Einstein equations are not the only one that undergo homogenization
under condition (\ref{conditionLT}).

Keeping this in mind let us consider a more general case,
when the source of the gravitational field is a fluid
with inhomogeneous pressure, viscosity, and acceleration.
Let us however keep spherically symmetry and $\omega_{ab} = 0$.
The most general form of spherical symmetric metric is

\begin{equation} 
{\rm d} s^2 = -{\rm e}^{A}{\rm d} t^2
+ {\rm e}^B {\rm d}  r^2 + R^2 {\rm d} \vartheta^2 + R^2 \sin^2 \vartheta {\rm d} \varphi^2. \end{equation}

In this case eq. (\ref{shearev}) reduces to

\begin{equation} 
 \dot{\sigma} 
=  \mathcal{D}  - \frac{1}{\sqrt{3}} \sigma^2 
- \frac{2}{3} \theta \sigma - \frac{1}{2 \sqrt{3}} (\rho -\overline{\rho}),
\end{equation}
where
$\mathcal{D} = 
\frac{\sqrt{3}}{2} {\rm e}^{-B} \left( \frac{1}{3} A'' - 
\frac{1}{6} A' B' + \frac{1}{6} A'^2 - \frac{1}{3} \frac{R'}{R} A' \right).$
Given that $ A' {\rm e}^{-B} = ({2\dot{E}})/({\dot{R}R'})$
the following conditions 

\begin{equation}
\frac{E'}{RR'} \ll \rho, \quad \frac{E}{R^2} \ll \rho, \quad
\frac{\dot{E}}{\dot{R}R} \ll \rho,
\label{conditions}
\end{equation}
are sufficient to ensure that $(2/3) \theta \sigma$ term is larger than others.
Although under these conditions shear vanishes, this is not a sufficient condition 
for homogenization, as there are solutions of the Einstein equations
that are shear-free but inhomogeneous (see \cite{CM94,MC94}). However, 
if in addition it is assumed that the rate of change of entropy 
is positive then, as shown by Ellis \cite{E71}, this implies

\begin{equation} \pi_{ab}  = - \lambda \sigma_{ab}, \end{equation}
where $\lambda >0$ is a viscosity coefficient.
So under the above condition 
vanishing shear implies vanishing viscosity and hence (\ref{conditions}) leads to homogenization.

Let us consider what these relations (\ref{conditions}) imply. 
They imply that the function $E$ is confined or small in comparison
to contribution from density. However, such a geometric explanation
does not provide physical insight.
In the case of vanishing heat flux, the conservation equations ($T^{a b}{}_{;b} = 0$)
reduce to

\begin{equation}  \frac{A'}{2} = \frac{-p' + \frac{2}{\sqrt{3}} ( \lambda \sigma)'
+2 \sqrt{3} \lambda \sigma R'/R 
}{\rho+p}. \end{equation}
Combining this with expression for $\mathcal{D}$ and demanding that 
all terms remain smaller than $\rho$ implies

\begin{equation}  \frac{\partial p}{\partial R} 
\sim \frac{\partial \lambda \sigma}{\partial R}
\ll \rho^{3/2}, \quad
\frac{\partial^2 p}{\partial R^2} 
\sim \frac{\partial^2 \lambda \sigma}{\partial R^2} 
\ll \rho^2, \end{equation}
which shows that although the gradients of pressure, shear and viscosity
 can tend to infinity when $t\to 0$
they cannot be unbound and must be smaller than $\rho^{3/2}$.
Given the fact that when matter is highly dense, its equation 
of state becomes degenerate (being weakly depend on temperature)
it might happen that pressure and its gradients
do not increase so rapidly as density as $t\to0$.
If this is indeed the case then conditions
(\ref{conditions}) may be more probable than it first appears.

Summarizing, it seems clear there are conditions 
in the early Universe that can lead to spontaneous homogenization
and which are not of measure zero. 
This is in contrast to study of decaying modes
in the Lema\^itre--Tolman or FLRW models.
The difference is that it
is not required that the model becomes homogeneous as $t\to \infty$,
but rather in the early stages of its evolution.
If the density is large and gradients of pressure and viscosity not extreme,
then relations (\ref{conditions}) ensure homogenization.
Thus as long as the density is of very large amplitude the homogenization proceeds.
It is interesting to notice that viscosity helps in keeping density at higher values.
If heat flow is zero then the conservation equation $T^{0 b}{}_{;b} = 0$ reduces to

\begin{equation} \dot{\rho} = - \theta (\rho+p) + \lambda \sigma^2.\end{equation}
This relation shows that in the presence of viscosity ($\lambda \sigma^2$) density decreases more
slowly than in the case of zero-viscosity.
In addition as shown in Ref. \cite{BL08} if gradients of pressure are present then 
the increase of density contrast in density waves 
(which can develop a shock when $p=0$)  is slower than in $p'=0$ case,
and eventually the gradients lead to acoustic oscillations.
Intuitively, one can imagine that if viscosity is present then
it is harder to induce such oscillations thus 
keeping the density contrast at lower values.
This all leaves an impression that dynamical processes of the early universe
could lead to homogenization.
Whether or not this took place in {\it our} Universe
remains to be examined by detailed numerical calculations.
Even if our Universe was not homogenized by 
some dynamical process, the main conclusion of this essays ramins unchanged:
a set of initial conditions for spontaneous homogenization of cosmological models
can form a set of non-zero measure.

\vspace{0.8cm}
\noindent {\it Acknowledgements}

\noindent KB would like to thank Paulina Wojciechowska for helpful comments and discussions concerning this essay. This research has been supported by the Polish--U.S. Fulbright Commission under grant PPLS/03/09.


\begin{thebibliography}{10}

\bibitem{EGS} 
J. Ehlers, P. Geren, and R.K. Sachs, {\it J. Math. Phys.}  {\bf 9}, 1344 (1968).

\bibitem{SME}
W.R. Stoeger, R. Maarteens, and G.F.R. Ellis, {\it Astrophys. J.} {\bf 443}, 1 (1995).

\bibitem{RS68} 
M.J. Rees and D.W. Sciama, {\it Nature\/} {\bf 217}, 511 (1968).

\bibitem{IS06}
K.T. Inoue and J. Silk, {\it Astrophys. J.} {\bf 648}, 23 (2006).

\bibitem{IS07}
K.T. Inoue and J. Silk, {\it Astrophys. J.} {\bf 664}, 650 (2007).


\bibitem{sn1}
A.G. Riess at al.,  {\it Astrophys. J.} {\bf 116}, 1009 (1998).

\bibitem{sn2}
S. Perlmutter et al., {\it Astrophys. J.} {\bf 517}, 565 ( 1999).

\bibitem{bao}
D.J. Eisenstein et al., {\it Astrophys. J.} {\bf 633}, 560 ( 2005).

\bibitem{cmb}
G. Hinshaw  et al., {\it Astrophys. J. Suppl.} {\bf 170}, 288 (2007).


\bibitem{W84}
R.M. Wald, {\it General Relativity} (The University of Chicago Press, Chicago, 1984).


\bibitem{E06}
G.F.R. Ellis, {\it Philosophy of Physics}, ed. J. Butterfield, J. Earman, D.M. Gabby, P. Rhagard, J. Woods.  (Elsevier, Amsterdam, 2006).

\bibitem{L90}
A.D. Linde, {\it Inflation and Quantum Cosmology} (Academic Press, Boston 1990).

\bibitem{VM05}
V. Mukhanov, \textit{Physical Foundations of Cosmology} (Cambridge University Press, Cambridge, 2005).

\bibitem{L08}
A. Linde, {\it Lect. Notes Phys.} {\bf 738}, 1 (2008).

\bibitem{GP89}
D.S. Goldwirth and T. Piran, {\it Phys. Rev.} {\bf D40}, 3263 (1989).

\bibitem{GP90}
D.S. Goldwirth and T. Piran, {\it Phys. Rev. Lett.} {\bf 64}, 2852 (1990).

\bibitem{GP92}
D.S. Goldwirth and T. Piran, {\it Phys. Rept.} {\bf 214}, 223  (1992).

\bibitem{GS92}
E. Calzetta and M. Sakellariadou, {\it Phys. Rev.} {\bf D45},  2802 (1992).

\bibitem{M68}
C.W. Misner, {\it Astrophys. J.} {\bf 151}, 431 (1968).

\bibitem{HE73}
S.W. Hawking and G.F.R. Ellis,
{\it The large scale structure of space-time} (Cambridge University Press, Cambridge, 1973).

\bibitem{E71}
G.F.R. Ellis, in {\it Proceedings of the XLVII Enrico Fermi Summer School},
 edited by R. K. Sachs, (Academic Press, New York, 1971) p 105;
reprinted with historical comments in {\it Gen. Rel. Grav.} {\bf 41}, 581 (2009).


\bibitem{CH73}
C.B. Collins and S.W. Hawking, {\it Astrophys. J.} {\bf 180}, 317 (1973).


\bibitem{B74}
W.B. Bonnor, {\it Mon. Not. Roy. Astr. Soc.} {\bf 167}, 55 (1974).

\bibitem{BT76}
W.B. Bonnor and N. Tomimura, {\it Mon. Not. Roy. Astr. Soc.} {\bf 175}, 85 (1976).


\bibitem{S77}
J. Silk, {\it Astron. Astrophys.} {\bf 59}, 53 (1977).

\bibitem{PK06}
J. Pleba\'nski and A. Krasi\'nski, {\it An introduction to general relativity 
and cosmology} (Cambridge University Press, Cambridge, 2006).

\bibitem{vEE96}
H. van Elst, and G.F.R. Ellis, {\it Class.
Quant. Grav.} {\bf 13}, 1099 (1996).


\bibitem{L33}
G. Lema\^{\i}tre, {\it Ann. Soc. Sci. Bruxelles} {\bf A53}, 51 (1933); English translation 
with historical comments in {\it Gen. Rel. Gravit.} {\bf 29}, 637 (1997).

\bibitem{T34}
R.C. Tolman, {\it Proc. Nat. Acad. Sci. USA}  {\bf 20}, 169 (1934); reprinted 
with historical comments in {\it Gen. Rel. Gravit.} {\bf 29}, 935 (1997).


\bibitem{S75}
P. Szekeres, {Commun. Math. Phys}. {\bf 41}, 55 (1975).

\bibitem{RS82}
S. R. Roy and J.P. Singh, {\it Indian J. Pure Appl. Math} {\bf 13}, 1285 (1982).

\bibitem{AK97}
A. Krasi\'nski, {\it Inhomogeneous Cosmological Models}  (Cambridge University Press, Cambridge, 1997).


\bibitem{CM94}
A.A. Coley and D.J. McManus, {\it Class. Quant. Grav.} {\bf 11}, 1261 (1994).

\bibitem{MC94}
D.J. McManus and A.A. Coley, {\it Class. Quant. Grav.} {\bf 11}, 2045 (1994).

\bibitem{BL08}
K. Bolejko and P.D. Lasky, {\it Mon. Not. R. Astron. Soc}., {\bf 391}, L59 (2008).



 \end{thebibliography}
\end{document}